\documentstyle[aps,prbbib,epsfig,floats]{revtex}
\begin{document}
\wideabs{
\title{Superconductivity in the t-J model}
\author{N. M. Plakida}
\address{
Joint Institute for Nuclear Research, Dubna, Russia\\
and\\
Max-Planck-Institut f\"ur Physik Komplexer Systeme, Dresden, Germany}
\date{17 October, 2002}
\maketitle
\begin{abstract}
A comparison of  microscopic  theories of superconductivity in
the limit of strong electron correlations is presented. The
results for the two-dimensional  t-J model obtained
within the projection technique for the Green functions in terms of
the Hubbard operators and the slave-fermion representation for the
RVB state are considered. It is argued that the latter approach
resulting in the  odd-symmetry p-wave pairing for fermions is inadequate.
\end{abstract}
 }


\section{Introduction}

A mechanism  of high-temperature superconductivity in cuprates is
still unresolved since  strong electron correlations  in
copper-oxygen planes prevent from applying   well established
methods of band structure calculations developed for conventional
metals. An important role of electron correlations in cuprates was
 initially stressed  by Anderson~\cite{Anderson87} who proposed to
consider them  within the framework of the Hubbard model or the
so-called $t$-$J$ model which follows from the Hubbard model in
the limit of strong correlations.
 To study these models, a lot of numerical work have
been  done~\cite{Scalapino95,Dagotto94} though the obtained
results are still controversial. For instance, a robust $d$-wave
pairing was observed for the $t$-$J$ model~\cite{Sorella01}, while
 a long-range order was not found in the original
Hubbard model.~\cite{Huang01}

 In analytical approache mostly a mean-field theory (MFA) was  applied
in studies of  the Hubbard model or the $t$-$J$ model.  The
resonating valence bond (RVB) state in the $t$-$J$ model was
proposed
 by Baskaran et al.~\cite{Baskaran87} where superconductivity was
 obtained as a result of spin correlations
induced by the superexchange interaction. Similar results were
found by Cyrot~\cite{Cyrot87} for superconducting pairing mediated
by superexchange interaction. To overcome the problem of strong
correlations in the $t$-$J$ model and nonfermionic commutation
relations for the physical electron operators a number of
auxiliary field representations was proposed (see,
e.g.,~\cite{Ruckenstein87}$^{-}$ \cite{Kochetov00}). However, in
these methods usually  a spin-charge separation is assumed for
spinon and holon fields   which violates rigorous restrictions
imposed by nonfermionic commutation relations as the "no double
occupancy" constraint which may  result in unphysical conclusions.

A reliable analytical approach to deal with strong correlations
in the Hubbard or the $t$-$J$ model is based on the  Green
function methods in terms of the Hubbard operators which
rigorously preserve the non-fermionic commutation
relations~\cite{Izyumov97}. Here we may mention our results based
on the Mori-type projection technique for the Green
functions~\cite{Plakida89}$^{-}$\cite{Plakida99} and the diagram
technique calculations by Izyumov et
al.~\cite{Izyumov91,Izyumov92}. These approaches enable one to go
beyond MFA by taking into account self-energy corrections. For
instance, a numerical solution of the Dyson equation
in~\cite{Plakida99} revealed a non Fermi-liquid behavior  in the
normal state at low doping and  the $d$-wave superconductivity
mediated by the exchange and spin-fluctuation pairing.  In the
recent paper~\cite{Plakida01} a microscopical theory of
superconductivity in CuO$_2$ layer within the effective two-band
$p$-$d$  Hubbard model in the strong correlation limit was
developed. It has been proved that  the MFA for the Hubbard model
results in the antiferromagnetic exchange $d$-wave pairing which
is equivalent to pairing observed in the $t$-$J$ model in MFA.

 In the present paper I compare the results
for the $t$-$J$ model obtained within the Green function (GF)
method in terms of the Hubbard
operators~\cite{Plakida89}$^-$\cite{Plakida99}
and by applying the
slave-fermion hard-boson representation~\cite{Feng94} for the RVB
state. It will be shown that the latter approach results in an
odd-symmetry $p$-wave pairing for the spinless fermions as in the
path-integral representation in~\cite{Kochetov00} which
contradicts to  known numerical and analytical calculations. It
casts doubts on the validity of spin-charge separation approach in
studying superconducting pairing in the $t$-$J$ model.

In the next Section we present briefly the results of the
projection technique for the GF~\cite{Plakida99} for the $t$-$J$
model. In Sec.\,3 superconducting pairing within the slave-fermion
representation for the Hubbard operators is considered. Results
and discussions are given in Sec.\,4. Concluding remarks are in
Sec.\,5.

\section{Green function method}
\subsection{Dyson Equation for the t-J Model }

 In the present Section we consider
superconducting pairing in the $t$-$J$ model by applying the GF
technique~\cite{Plakida89}$^{-}$\cite{Plakida99}. The $t$-$J$
model in the standard notation~\cite{Anderson87,Zhang88} reads:
\begin{equation}
H_{t-J}= - t \sum_{i \neq j, \sigma} \tilde a_{i\sigma}^{+}
\tilde a_{j\sigma}
 +  J\sum_{\langle ij \rangle}
( \mbox{\bf S}_{i}\mbox{\bf S}_{j} - {\frac {1}{4}} n_{i} n_{j}),
\label{1}
\end{equation}
where $\tilde a_{i\sigma}^{+}=a_{i\sigma}^{+}(1-n_{i-\sigma})$ are
 projected  operators for physical electrons,
  $\, n_i = \sum_{\sigma}\tilde a_{i\sigma}^{+} \tilde a_{i\sigma}$ is the number
operator  and $S_{i}^{\alpha} = (1/2)\sum_{s,s'}\tilde
a_{is}^{+}\sigma^{\alpha}_{s,s'}\tilde a_{is'}$ are  spin-1/2
operators. Here $t$ is   effective transfer integral and $J$ is
 antiferromagnetic  exchange energy   for a pair of nearest
neighbor spins, $\; \langle ij\rangle, \; i>j \;$.

To take into account on a rigorous basis the projected character
of electron operators  we employ the Hubbard operator  technique.
The Hubbard operators (HO)  are defined as $\,
X_{i}^{\alpha\beta}=|i,\alpha\rangle\langle i,\beta| \,$ for three
possible states $|i,\alpha\rangle\,$  at a lattice site $i$:
 for an empty site $\,|i,0\rangle \;$ and
 for a singly occupied site $\;|i,\sigma\rangle \,$  by an electron
with spin $\sigma/2$ ($\sigma=\pm 1 , \; \bar{\sigma} = -
\sigma$). They obey the completeness relation
\begin{equation}
 X_{i}^{00}+\sum_{\sigma}X_{i}^{\sigma\sigma} = 1,
\label{3}
\end{equation}
which rigorously preserves the constraint of no double occupancy.
The spin and density operators in Eq.(\ref{1}) are expressed by
HO as
\begin{equation}
S_{i}^{\sigma}=X_{i}^{\sigma\bar{\sigma}},\;\;\;
S_{i}^{z}=\frac{1}{2}\sum_{\sigma}\sigma
X_{i}^{\sigma\sigma},\;\;\;
n_{i}=\sum_{\sigma}X_{i}^{\sigma\sigma}. \label{4}
\end{equation}
The HO obey the following multiplication rule $\,
X_{i}^{\alpha\beta}X_{i}^{\gamma\delta}= \delta_{\beta\gamma}
X_{i}^{\alpha\delta}\, $ and commutation relations
\begin{equation}
\left[X_{i}^{\alpha\beta}, X_{j}^{\gamma\delta}\right]_{\pm}=
\delta_{ij}\left(\delta_{\beta\gamma}X_{i}^{\alpha\delta}\pm
\delta_{\delta\alpha}X_{i}^{\gamma\beta}\right),
 \label{6}
\end{equation}
where the upper sign stands for the  the Fermi-like HO (as, e.~g.,
$X_{i}^{0\sigma}$) and the lower sign for the Bose-like  operators
(as the spin and number operators in Eq.~(\ref{4}) ).

By using the Hubbard operator representation  we write the
Hamiltonian of the $t$-$J$ model (\ref{1}) in a more general form:
\begin{eqnarray}
H_{t-J}&=& - \sum_{i \neq j, \sigma}t_{ij}X_{i}^{\sigma
0}X_{j}^{0\sigma}
 - \mu \sum_{i \sigma} X_{i}^{\sigma \sigma}
 \nonumber\\
 & +& \frac{1}{4} \sum_{i \neq j, \sigma} J_{ij}
\left(X_i^{\sigma\bar{\sigma}}X_j^{\bar{\sigma}\sigma}  -
   X_i^{\sigma\sigma}X_j^{\bar{\sigma}\bar{\sigma}}\right).
\label{9}
\end{eqnarray}
The  electron hopping  energy for the nearest neighbors, $t_{ij}=
t $, and the second neighbors, $t_{ij}=  t^{'}$, on a 2D square
lattice, and the exchange interaction  $J_{ij}= J$ for the nearest
neighbors~\cite{exchange} can be considered as independent
parameters if, starting from  a more realistic for copper oxides
three-band $p$-$d$ model~\cite{Emery87}, we reduce it to the
$t$-$J$ model~\cite{Zhang88}. In that case the parameters $\; t,
t'$ and $J$ can be evaluated in terms of the original parameters
of the $p$-$d$ model~(see, e.g.,~\cite{Feiner96,Yushankhai97}).
We introduced also the  chemical potential $\mu$ which can be
calculated from the equation for the average number of electrons
\begin{equation}
n= \langle n_i \rangle = \sum_{\sigma} \langle X_{i}^{\sigma
\sigma} \rangle . \label{7}
\end{equation}

To discuss the superconducting pairing within the model (\ref{9})
we introduce the Nambu notation for HO:
\[\Psi_{i \sigma} = \left( \begin{array}{c} X_i^{0 \sigma } \\
X_i^{ \bar{\sigma} 0} \end{array} \right),\qquad \Psi_{i
\sigma}^+ = \left( X_i^{ \sigma 0} \; X_i^{0 \bar{\sigma} }
\right)\; ,
\]
and consider the  matrix GF
\begin{equation}
\hat{G}_{i j,\sigma} (t-t') =
  \langle  \langle \Psi_{i \sigma}(t) | \Psi_{j \sigma}^+ (t') \rangle\rangle
  = \left( \begin{array}{cc}
     G\sb{ij\sigma}^{11}& G\sb{ij\sigma}^{12}\\
    G\sb{ij\sigma}^{21} &  G\sb{ij\sigma}^{22}
  \end{array} \right)
\label{10}
\end{equation}
where Zubarev's notation for the anticommutator GF is
used~\cite{Zubarev60}.

By differentiating the GF (\ref{10}) over the time $t$ we get for
the Fourier component the following equation
\begin{equation}
\omega \hat {G}_{ij\sigma} (\omega) = \delta_{ij} \hat
{Q_{\sigma}} + \langle\!\langle \hat{Z}_{i\sigma} \mid  \Psi_{j
\sigma}\rangle\!\rangle_{\omega},
 \label{12}
\end{equation}
where $\hat {Z}_{i\sigma} = [\Psi_{i\sigma}, H] \; $,
 $\hat {Q_{\sigma}}=
\left( \begin{array}{cc} Q_{\sigma}& 0 \\
                                  0&  Q_{\bar \sigma} \end{array} \right) $
with $ Q_{\sigma} = \langle  X_i^{ 0 0} + X_i^{ \sigma \sigma }
\rangle$. Since we consider a spin-singlet state the correlation
function $\, Q_{\sigma}= Q =1-n/2 \; $   depends only on the
 average number of electrons~(\ref{7}).

Now, we project the many--particle GF in (\ref{12}) on the
original single--electron GF
\begin{eqnarray}
\langle\!\langle \hat {Z}_{i\sigma} \mid  \Psi^+_{j\sigma}
\rangle\!\rangle &=& \sum_l \hat{E}_{il\sigma} \langle\!\langle
\Psi_{l\sigma}\mid \Psi^+_{j\sigma}\rangle\!\rangle
\nonumber \\
 &+& \langle\!\langle \hat {Z}_{i\sigma}^{(irr)} \mid  \Psi^+_{j\sigma}
\rangle\!\rangle,
\label{13}
\end{eqnarray}
where the irreducible ({\em irr}) part  of the many--particle
operator $\hat {Z}_{i\sigma}$ is defined by the equation
\[
\langle \{ \hat{Z}^{(irr)}_{i\sigma}, \Psi^{+}_{j\sigma} \}
\rangle = \langle \{ \hat{Z}_{i\sigma} -\sum_l
\hat{E}_{il\sigma}\Psi_{l\sigma},\, \Psi^{+}_{j\sigma} \}
 \rangle =0 ,
 \]
which results in the definition of the frequency matrix
\begin{equation}
\hat {E}_{ij\sigma} = \langle \{ [\Psi_{i\sigma}, H],
\Psi^+_{j\sigma}\} \rangle \; {Q}^{-1} \; .
\label{17}
\end{equation}
Now we can introduce the zero--order GF in the generalized MFA
which is given by the frequency matrix (\ref{17})
\begin{equation}
\hat {G}^0_{ij\sigma}(\omega) = Q \{\omega \hat \tau_0
\delta_{ij} - \hat {E}_{ij\sigma} \}^{-1} . \label{23}
\end{equation}

To derive the Dyson equation for the single-electron GF~(\ref{10})
we  write down an equation of motion for the   irreducible part of
the GF in~(\ref{13}) with respect to the second time $t'$ for the
right--hand side operator $\Psi^+_{j\sigma} (t')$. Then performing
the same projection procedure as in Eq.~(\ref{13}) we obtain the
Dyson equation for the GF  in the form
\begin{equation}
\hat {G}_{ij\sigma} (\omega) = \hat {G}^0_{ij\sigma}(\omega) +
\sum_{kl} \hat{G}^0_{ik\sigma} (\omega)\
\hat{\Sigma}_{kl\sigma}(\omega) \ \hat{G}_{lj\sigma} (\omega) ,
\label{25}
\end{equation}
where the self--energy operator $\hat{\Sigma}_{kl\sigma} (\omega)$
is defined by the equation
\begin{equation}
\hat{T}_{ij\sigma} (\omega) = \hat \Sigma_{ij\sigma} (\omega) +
\sum_{kl} \hat {\Sigma}_{ik\sigma}(\omega) \ \hat
{G}^0_{kl\sigma}(\omega) \ \hat{T}_{lj\sigma}(\omega) \; .
\label{26}
\end{equation}
Here the scattering matrix is given by
\begin{equation}
\hat {T}_{ij\sigma} (\omega) = Q^{-1} \langle\!\langle
\hat{Z}^{(irr)}_{i\sigma} \mid \hat {Z}^{(irr)^+}_{j\sigma}
\rangle\!\rangle_{\omega} \ Q^{-1} .
 \label{27}
\end{equation}
From Eq.~(\ref{26}) it follows that the self-energy operator is
given by the irreducible part of the scattering matrix (\ref{27})
that  has no parts connected by the single zero-order GF
(\ref{23}):
\begin{equation}
\hat {\Sigma}_{ij\sigma} (\omega) = Q^{-1} \langle\!\langle
\hat{Z}^{(irr)}_{i\sigma} \mid \hat {Z}^{(irr)^+}_{j\sigma}
\rangle\!\rangle^{(irr)}_{\omega}  \;
 Q^{-1}   \; .
\label{28}
\end{equation}
Equations~(\ref{23}), (\ref{25}) and (\ref{28}) give  an exact
representation for the single--electron GF (\ref{10}). However, to
solve the self-consistent system of equations  one has to
introduce an approximation for the many--particle GF in the
self-energy matrix (\ref{28}) which describes inelastic
scattering of electrons on spin and charge fluctuations.
\subsection{ Self-Consistent  Equations}
Here we derive  a self-consistent system of equations in MFA. To
calculate the frequency matrix~(\ref{17}) we use the equation of
motion for the HO:
\begin{eqnarray}
\left( i\frac{d}{dt} +\mu \right) X_{i}^{0\sigma}& = &
 -  \sum_{l,\sigma'} t_{il} B_{i \sigma \sigma'} X_{l}^{0\sigma'}
\nonumber \\
&+& \frac {1}{2} \sum_{l,\sigma'} J_{il} (B_{l \sigma
\sigma'}-\delta_{\sigma \sigma'}) X_{i}^{0\sigma'},
 \label{18}
\end{eqnarray}
where we introduced the   operator
\begin{eqnarray}
 B_{i\sigma\sigma'} & = &
     (X^{00}_{i} + X^{\sigma\sigma}_{i})\delta_{\sigma'\sigma}
  +   X^{\bar{\sigma}\sigma}_{i}\delta_{\sigma' \bar{\sigma}}
\nonumber \\
& = & (1- \frac{1}{2} n_i +  S^z_i) \delta_{\sigma'{\sigma}}
 + S^{\bar{\sigma}}_{i} \delta_{\sigma' \bar{\sigma}} .
\label{18a}
\end{eqnarray}
The Bose-like operator (\ref{18a}) describes electron scattering
on spin and charge fluctuations caused by   the kinematic
interaction (the first term in~(\ref{18})) and by  the exchange
spin-spin interaction (the second term in~(\ref{18})).

 By performing commutations
in~(\ref{17}) we get for the normal and the anomalous parts of the
frequency matrix:
\begin{eqnarray}
 {E}_{ij\sigma}^{11} &=& -\mu \delta_{ij} + \delta_{ij} \sum_{l}
    \{ t_{il}  \langle X_{i}^{\sigma 0}X_{l}^{0\sigma} \rangle /Q
\nonumber \\
&+ &\frac{1}{2} J_{il} (Q - 1  + \chi_{il}^{cs} /Q) \}
\nonumber \\
& - & t_{ij} ( Q+  \chi_{ij}^{cs}/ Q )
 - \frac{1}{2} J_{ij} \langle X_{j}^{\sigma 0}X_{i}^{0\sigma} \rangle / Q
\label{20}
\end{eqnarray}
\begin{eqnarray}
 {E}_{ij\sigma}^{12} = \Delta_{ij\sigma} &=& \delta_{ij}  \sum_{l} t_{il}
  \langle X_{i}^{0\bar{\sigma}}X_{l}^{0\sigma}  -
  X_{i}^{0\sigma} X_{l}^{0\bar{\sigma}} \rangle /Q
\nonumber \\
 &-& \frac{1}{2} J_{ij} \langle X_{i}^{0\bar{\sigma}}X_{j}^{0\sigma}
-  X_{i}^{0\sigma} X_{j}^{0\bar{\sigma}} \rangle /Q \; .
\label{21}
\end{eqnarray}
Here  in calculation of the correlation function for the normal
component of the frequency matrix:
\begin{eqnarray}
&&\sum_{\sigma'} \langle  B_{i\sigma\sigma'} B_{j\sigma'\sigma}
\rangle =
 \nonumber \\
  &&\langle (1- \frac{1}{2}n_i +  S^z_i) (1 - \frac{1}{2}n_j
+  S^z_j) + S^{\sigma}_i S^{\bar\sigma}_j \rangle
 \nonumber \\
 &=& \langle(
1 - \frac{1}{2}n_i)( 1 - \frac{1}{2}n_j) \rangle + \langle {\bf
S}_i {\bf S}_j \rangle  = \;  Q^2 \;  + \; \chi_{ij}^{cs} \; ,
 \nonumber
\end{eqnarray}
we introduce the charge- and spin-fluctuation correlation
functions
\[
\chi_{ij}^{cs} = \frac{1}{4} \langle \delta n_{i} \delta
n_{j}\rangle + \langle {\bf S_{i}}{\bf S_{j}}\rangle  \; ,
\]
with $\delta n_{i}= n_{i}-\langle
n_{i}\rangle$. Further we neglect the charge fluctuations,
 $\, \langle \delta n_{i} \delta  n_{j}\rangle \simeq 0 \,$,
 but take into account spin
correlations given by  spin correlation functions for the nearest
($\chi_{1s} $) and the next-nearest ($ \chi_{2s}  $) neighbor
lattice sites:
\begin{equation}
 \chi_{1s}=\langle {\bf S}_{i} {\bf S}_{i+a_{1}}  \rangle  \; , \quad
 \chi_{2s}=\langle {\bf S}_{i}{\bf S}_{i+a_{2}}  \rangle ,
\label{34}
\end{equation}
where $a_{1}=(\pm a_{x}, \pm a_{y})$ and  $a_{2}=\pm (a_{x} \pm
a_{y})$. In a paramagnetic state assumed here they  depend only on
the distance between the lattice sites.

In the  ${\bf k}$-representation for the GF
\[
G^{\alpha\beta}_{\sigma} ({\bf k},\omega) = \sum_{j}
G^{\alpha\beta}_{oj\sigma} (\omega) \;
\mbox{e}^{-i{\mbox{\small\bf kj}}} \; ,
\]
we get for the zero-order GF (\ref{23}):
\begin{eqnarray}
\hat G^{(0)}_{\sigma} ({\bf k},\omega)&= & Q \, \{
\omega\hat\tau_{0} - (\varepsilon_{\bf k} -\tilde \mu) \hat
\tau_{3}-
 \Delta_{\bf k}^{\sigma} \hat \tau_{1} \}^{-1}
 \nonumber \\
 &=& Q \,\frac{\omega \hat\tau_{0}+(\varepsilon_{\bf k} -\tilde \mu)
 \hat \tau_{3}+\Delta_{\bf k}^{\sigma}\hat \tau_{1} }
 {\omega^{2}-(\varepsilon_{\bf k} -\tilde \mu)^{2}-
 \mid\Delta_{\bf k}^{\sigma}\mid^{2} },
 \label{30}
\end{eqnarray}
where $\hat \tau_{0}, \; \hat \tau_{1}, \; \hat \tau_{3} $ are the
 Pauli matrices.
  The quasiparticle  energy   $\, \varepsilon_{\bf k}$ and
 the renormalized chemical potential  $\, \tilde \mu = \mu -\delta \mu $
  in the MFA   are defined  by the frequency matrix~(\ref{20})
\begin{eqnarray}
 \varepsilon_{\bf k} &=& - \tilde  t({\bf k})
 - \frac{2J}{N} \sum_{\bf q} \gamma({\bf k-q}) N_{{\bf q} \sigma},
\label{31}\\
 \delta \mu & = & \frac{1}{N}\sum_{\bf q}\, t({\bf q})N_{{\bf q}\sigma} -
 2J({n}/{2} - {\chi_{1s} }/{ Q}) \;,
\label{32}
\end{eqnarray}
where $J({\bf q}) = 4J\gamma({\bf q}) ,\;
  t({\bf k})= 4t\gamma({\bf k}) +4t'\gamma'({\bf k}) \; $
 with $\;
\gamma({\bf k})=(1/2)(\cos a_{x}q_{x}+\cos a_{y} q_{y}) \; $,
 $\; \gamma'({\bf k})=\cos a_{x}q_{x}\cos a_{y}q_{y}\,$, while the
 renormalized hopping integral is given by
\begin{eqnarray}
\tilde  t({\bf k}) & = &4 t\,  \gamma({\bf k})\,Q \,(1+
\chi_{1s}/Q^2)
 \nonumber \\
& + & 4 t' \, \gamma'({\bf k})\,Q\, (1+ \chi_{2s}/Q^2).
\label{32a}
\end{eqnarray}
The  average number of electrons in
Eqs.~(\ref{31}),~(\ref{32}) in the ${\bf k}$-representation is
written in the form:
\[n_{{\bf k},\sigma}=  \langle X_{{\bf k}}^{\sigma 0}  X_{{\bf
k}}^{0 \sigma}  \rangle
  = Q  N_{{\bf k}\sigma}  \; .
\]
It should be pointed out that the renormalization of the hopping
parameter~(\ref{32a}) caused  by the spin correlation functions
(\ref{34}) are essential at low doping when short-range
antiferromagnetic correlations are strong. For instance,  for hole
doping $\, \delta = 1- n \simeq 0.05\,$ and $\, Q =(1+ \delta)/2
\simeq 0.53 \,$  the correlation functions are estimated
as~\cite{Plakida99}: $\,\chi_{1s} \simeq -0.3, \, \chi_{2s}
\simeq 0.2 ,\,$  which results in complete suppression of the
nearest neighbors hopping, while the next-neighbor hopping is
quite large:
\[
 t_{eff} = t\; Q [1 + \chi_{1s}/ Q^2] \simeq 0\, ,
 \]
\[
{t'}_{eff}=  t' \; Q [1 + \chi_{2s}/ Q^2 ] \simeq 0.9 \, t'\, .
\]
For large doping the antiferromagnetic correlations are suppressed
and the nearest neighbor hopping prevails: $\, t_{eff}/ {t'}_{eff}
\simeq t/t' \gg 1 \,$.

The superconducting  gap $\, \Delta_{{\bf k}}^{\sigma}\,$ in
Eq.~(\ref{30}) is defined by the anomalous component of the the
frequency matrix~(\ref{21}):
 \begin{equation}
\Delta_{{\bf k}}^{\sigma} =   \frac{2}{N Q} \sum_{{\bf q}} \; [\,
t({\bf q})- \frac {1}{2} J({\bf k-q} \,] \;
 \langle X_{{\bf -q}}^{0 \bar{\sigma}}  X_{{\bf q}}^{0 \sigma} \rangle ,
\label{33}
\end{equation}
There are two contributions in Eq.~(\ref{33}) given by the ${\bf
k}$-independent kinematic interaction $t({\bf q})$ and the
exchange interaction $J({\bf k-q})$. The kinematic interaction
 gives  no contribution to the $d$-wave pairing in MFA,
Eq.~(\ref{33}) (see~\cite{Plakida89}), and we disregard it in the
following equations~\cite{Zaitsev87}. The anomalous correlation
function in Eq.~(\ref{33}) can be easily calculated from the
anomalous part of the GF~(\ref{30}):
\begin{equation}
\langle X_{{\bf -q}}^{0 \bar{\sigma}}  X_{{\bf q}}^{0 \sigma}
\rangle = - Q\, \frac{\Delta_{{\bf q}}^{\sigma}}{2 E_{\bf q}}
\tanh\frac{E_{\bf q}}{2T}\, ,
 \label{33a}\end{equation}
which results  in the BCS-type equation for the gap function:
\begin{equation}
\Delta_{\bf k}^{\sigma} =  \frac{1}{N} \sum_{\bf q} \, J({\bf
k-q}) \frac{\Delta_{{\bf q}}^{\sigma}}{2 E_{\bf q}}
\tanh\frac{E_{\bf q}}{2T},
 \label{35}
\end{equation}
where $\, E_{k}=[\,( \varepsilon_{\bf k}
  - \tilde{\mu})^{2} + \mid \Delta_{\bf k}^{\sigma}\mid^{2}\,]^{1/2}\; $
is the quasiparticle energy in the superconducting  state. As was
proved in~\cite{Plakida01}, the retardation effects for the
exchange interaction are negligible and therefore there is no
restriction in integration over the energy in Eq.~(\ref{35}). It
means that all electrons  in the conduction band participate in
the superconducting pairing contrary to the BCS equation for the
electron-phonon model where the energy integration and pairing are
restricted to a narrow energy shell of the order of the phonon
energy close to the Fermi energy.

The equation~(\ref{35}) is identical to  the results in MFA of the
diagram technique~\cite{Izyumov91}, while the gap equation
obtained in~\cite{Onufrieva96} has an additional factor
$Q=(1-n/2)\,$ which is spurious.  This factor  appears if we apply
a simple decoupling procedure
 in the equation of motion~(\ref{18}) for GF instead of the projection
technique given by Eq.~(\ref{17}). Writing the bosonic operators
in the exchange interaction as a product of two fermionic
operators:
 $\, X_i^{\bar\sigma {\sigma}} = X_{i}^{\bar\sigma 0} X_{i}^{0\sigma}\, $
and performing decoupling of the fermionic operators:
\begin{eqnarray}
&&\langle\langle X_{l}^{\bar\sigma 0} X_{l}^{0 \sigma} X_{i}^{0
\bar \sigma} - X_{l}^{\bar\sigma 0}X_{l}^{0 \bar\sigma}
X_{i}^{0\sigma}\mid X_{j}^{\sigma 0} \rangle\rangle_{\omega}
\simeq
 \nonumber \\
  && \langle X_{l}^{0\sigma}X_{i}^{0 \bar\sigma} -
 X_{l}^{0\bar\sigma}X_{i}^{0\sigma}\rangle \,
 \langle\langle X_{l}^{ \bar\sigma 0}\mid
 X_{j}^{\sigma0}\rangle\rangle_{\omega}\, ,
 \label{11a}
\end{eqnarray}
we obtain the same equation~(\ref{35}) for the gap function but
with the additional $Q$-factor at the right hand side. In the
decoupling we miss the normalization factor for the  correlation
functions in the denominator of the frequency matrix,
Eq.~(\ref{17}), which cancels out with $Q$ in the numerator of the
corresponding GF~(\ref{30}). So, a rigorous way to apply  MFA with a proper
account of projected character of HO is to use the projection
technique as discussed above.

 The self-energy contribution~(\ref{28}) in  the second order of the
the kinematic interaction is considered in~\cite{Plakida99} while
it is omitted in ~\cite{Izyumov91}. As discussed
in~\cite{Plakida99,Plakida01}, it mediates the spin-fluctuation
pairing and results in finite life-time effects for the
quasiparticle spectrum giving rise to an incoherent contribution
to the single-particle density of states. Here we shall not
discuss further these self-energy effects since for comparison the
GF approach with the slave-fermion technique it will be sufficient
to consider only MFA for the gap equation~(\ref{35}).

\section{Slave-fermion approach}
\subsection{Slave-fermion representation}
A number of auxiliary field  representations has been proposed so
far (see, e.g.,~\cite{Ruckenstein87}$^{-}$\cite{Kochetov00}). In
the slave-boson
method~\cite{Ruckenstein87}$^{-}$\cite{Suzumura88} the projected
electron operator is expressed as a product of the auxiliary Bose
field for the charge degree of freedom (holon) and the Fermi
field for the spin degree of freedom (spinon). The main problem
in this approach is the so called constraint imposed by projected
character of electronic operators in the $t$-$J$ model which
prohibits double occupancy of any lattice site. To treat the
constraint a site-dependent  Lagrange multiplier is introduced.
However, to solve the problem the  MFA is usually applied and the
Lagrange multiplier is taken to be independent on the lattice site
so the local constraint is replaced by a global one with
uncontrollable consequences. In the slave-fermion method the
charge degree of freedom is represented by spinless fermion
operators, while to describe the spin degree of freedom  the Bose
field (Schwinger
bosons~\cite{Yoshioka89}$^{-}$\cite{Khaliullin93} or spin
operators~\cite{Khaliullin90}$^{-}$\cite{Feng94}) are used. The
Schwinger boson representation though being physically meaningful
for the Heisenberg model~\cite{Arovas88,Auerbach88} gives poor
results for the doped case: the antiferromagnetic ground state
persists up to very high doping and it does not reproduce the
large Fermi surface as in the slave-boson method. In the slave
fermion and spin operator representation the magnetic properties
of the model are described in a more reliable way (see,
e.g.,~\cite{Wang93,Feng98,Yuan01}).

Below we consider the slave-fermion hard-core $(\rm CP^{1})$
boson representation
proposed in~\cite{Feng94} and later on employed
in~\cite{Feng97}$^{-}$\cite{Yuan01} to investigate different
physical properties of cuprates within the $t$-$J$ model. It has
some advantage since the constraint of no double occupancy can be
fulfilled without introducing the Lagrange multiplier. To
decouple the charge and spin degrees of freedom for  physical
electrons the HO in the theory~\cite{Feng94} is represented as a
product of a spinless fermion $\, h^{+}_i \,$ for the charge
degree of freedom (holon) and a hard-core boson $\, b_{i\sigma}
\,$ for the spin degree of freedom (spinon):
\begin{equation}
 X_{i}^{0\sigma}= h^{+}_i \, b_{i\sigma}, \quad
  X_{i}^{\sigma 0}= h^{}_i \, b_{i\sigma}^{+},
 \label{s1}
\end{equation}
which have the following commutation relations:
\[
 h^{+}_i h_j + h_j h^{+}_i = \delta_{i,j}, \quad
  b_{i\sigma} b_{j\sigma}^{+} -b_{j\sigma}^{+} b_{i\sigma}=
  \delta_{i,j}(1-2 b_{i\sigma}^{+} b_{i\sigma}).
\]
The hard-core bosons are  Pauli operators which commute on
different lattice sites and anticommute on the same lattice site
prohibiting  double occupancy. The Pauli operators  can be also
represented by the spin-lowering $S_{i}^{-}$ and spin-raising
$S_{i}^{+}$ operators for spin-$1/2$ :
\[
 b_{i\uparrow}^{+}=   S_i^{+} , \;
  b_{i\uparrow} =  S_i^{-} ,
 \quad  {\rm or} \; \quad
b_{i\sigma}^+ = S_i^{\sigma}, \; b_{i\sigma} = S_i^{\bar \sigma}.
\]
The on-site electron local constraint
\begin{eqnarray}
\sum_{\sigma}\, X_{i}^{\sigma 0}X_{i}^{0\sigma }&=&
\sum_{\sigma}\,  X_{i}^{\sigma \sigma}=
 h^{}_i h^{+}_i \sum_{\sigma} \, b_{i\sigma}^{+} \, b_{i\sigma}
\nonumber \\
  &= &  h^{}_i h^{+}_i = 1-  h^{+}_i h^{}_i \le 1
\label{s4}
\end{eqnarray}
is satisfied here since for the Pauli operators at any lattice
site we have the equation
\begin{equation}
\sum_{\sigma} \, b_{i\sigma}^{+} \, b_{i\sigma} =
  S_i^{+} S_i^{-} + S_i^{-}S_i^{+} = 1,
\label{s4a}
\end{equation}
and  the spinless holon number $\, n_i^{(h)} = h^{+}_i h^{}_i$
can be equal to 1 or 0.

However, the spin-charge separation imposed by the
representation~(\ref{s1}) results in  extra degrees of freedom: a
spin $1/2$ is assigned to any lattice site including an empty
site, while in the HO representation, Eq.~(\ref{3}), we have only
3 states: an empty state and a filled state with  spin $\pm 1/2$.
To cure this defect one should introduce a projection operator to
exclude the unphysical states~\cite{Feng94} (The same applies to
other slave-fermion spin operator representations as
in~\cite{Khaliullin90}$^{-}$\cite{Wang94}). Otherwise the
commutation relations for the original HO, Eq.~(\ref{6}), and
their representation, Eq.~(\ref{s1}), will give different results.
 For instance, for the physical electrons which are described by HO
we have
\begin{eqnarray}
&&\{X_{i}^{0\sigma},\, X_{i}^{\sigma 0}\}=  X_{i}^{00} +
X_{i}^{\sigma\sigma}= 1- X_{i}^{\bar\sigma \bar\sigma},
 \nonumber \\
{\rm and}\; & &\; \sum_{\sigma}\,\langle \{X_{i}^{0\sigma},\,
X_{i}^{\sigma 0}\}\rangle = 2 - \langle n_i\rangle =1 + \delta ,
\label{s5a}
\end{eqnarray}
where the hole doping concentration $\, \delta = 1- n \,$. If we
use the representation~(\ref{s1}) then we can write the
commutation relations as
\begin{eqnarray}
\{X_{i}^{0\sigma},\, X_{i}^{\sigma 0}\}&=& \{  h^{+}_i \,
b_{i\sigma}\, ,  h^{}_i \, b_{i\sigma}^{+} \}
 \nonumber \\
&=&
 h^{+}_i \, h_i + (1- 2\, h^{+}_i \, h_i ) b_{i\sigma}^{+}b_{i\sigma}\,
 \nonumber \\
{\rm and}&&  \sum_{\sigma}\, \{  h^{+}_i \, b_{i\sigma}\, , h^{}_i
\, b_{i\sigma}^{+} \}=  1,
 \label{s5b}
\end{eqnarray}
where we have used Eq.~(\ref{s4a}).
 For the average value in
Eq.~(\ref{s5b}) we get, respectively,
\begin{eqnarray}
 \sum_{\sigma}\, \langle \{  h^{+}_i \, b_{i\sigma}\, ,
h^{}_i \, b_{i\sigma}^{+} \}\rangle  & = & 1\, ,
  \label{s5d}
\end{eqnarray}
which contradicts to the rigorous result for HO, Eq.~(\ref{s5a}).
For the average number of electrons in the
representation~(\ref{s1}) by using the definition~(\ref{7}) we can
write
\begin{eqnarray}
  n & = & \langle n_i \rangle = \sum_{\sigma} \langle X_{i}^{\sigma0}
X_{i}^{0\sigma} \rangle
 \nonumber \\
& = & \sum_{\sigma} \langle  h^{}_i
 b_{i\sigma}^{+}h^{+}_i b_{i\sigma} \rangle =
\langle  h_i h^{+}_i \rangle = 1- \delta,
 \label{s5c}
\end{eqnarray}
which coincides with the definition~(\ref{7}) if we take for the
hole number operator the following definition: $ X_{i}^{00}
 = h^{+}_i h_i  $.
However, this definition is not unique. For instance we can write:
 $\, X_{i}^{00}= X_{i}^{0\sigma}X_{i}^{\sigma0}=
 X_{i}^{0\bar\sigma}X_{i}^{\bar\sigma0}\, $, which results in the
 equation: $\, X_{i}^{00} = (1/2)h^{+}_i h_i \, $ if we use the
 representation~(\ref{s1}) and the condition(\ref{s4a}).
So the double counting of  empty sites results in controversial
equation for an average number of electrons which is therefore
valid only with accuracy of  $\pm \delta $.

\subsection{Mean-field approximation}

 Let us consider the resonating valence bond (RVB) state in the original
 Hamiltonian~(\ref{9}) as proposed by Baskaran et al.~\cite{Baskaran87}.
 For this we should write the Bose-like
spin operators in the exchange energy in $H_{t-J}$~(\ref{9}) as a
product of two single-particle Fermi-like operators:
 $\, X_i^{\sigma\bar{\sigma}} = X_{i}^{\sigma0} X_{i}^{0\bar\sigma}\, $
 and  introduce the singlet operators
\begin{equation}
   b_{ij}^{\uparrow} \equiv  b_{ij}= \frac{1}{\sqrt{2}}( X_i^{0+}X_j^{0-} -
   X_i^{0-}X_j^{0+} )\, .
\label{m1}
\end{equation}
Then using MFA for the singlet operators  in the exchange
interaction of the $t$-$J$ model we get the RVB effective
Hamiltonian:
\begin{eqnarray}
H_{J} &=&\frac{1}{2} \sum_{i \neq j} J_{ij} ( X_i^{+-}X_j^{-+} -
   X_i^{++}X_j^{--})
=  - \frac{1}{2} \sum_{i \neq j} J_{ij}  b^{+}_{ij} b_{ij}
\nonumber \\
& \simeq & - \frac{1}{2} \sum_{i \neq j} J_{ij} \,
\left( B^{+}_{ij} b_{ij} +
 b^{+}_{ij} B_{ij}  -
 |B_{ij}|^2 \right),
\label{m2}
\end{eqnarray}
where we introduced the RVB order parameter:
\begin{equation}
 B_{ij}^{(\uparrow)} \equiv B_{ij} =   \langle b_{ij} \rangle =
 \frac{1}{\sqrt{2}}\langle X_i^{0+}X_j^{0-} -
   X_i^{0-}X_j^{0+}\rangle .
\label{m3}
\end{equation}
Here we should point out that in MFA~(\ref{m2}) a decoupling of
the Hubbard operators on the same lattice site is used:
\begin{eqnarray}
X_i^{+-}X_j^{-+}& = & X_i^{+0}X_i^{0-} X_j^{-0}X_j^{0+}
 \nonumber \\
&\Rightarrow & \;  \langle X_i^{+0}X_j^{-0} \rangle \;
X_j^{0+}X_i^{0-} \, , \label{m3a}
\end{eqnarray}
which is not unique and results in uncontrollable approximation
since the Hubbard operators obey the multiplication rule:
$\,X_{i}^{\alpha\beta}X_{i}^{\beta\gamma} = X_{i}^{\alpha\gamma}
\,$, and any intermediate state $\beta$ can be used in the
decoupling.

To obtain a self-consistent equation for the order parameter we
assume a spin-charge separation as it is usually done in the
slave-particle methods by applying a decoupling of spinon and
holon degrees of freedom introduced in Eq.~(\ref{s1}):
\begin{eqnarray}
   B_{ij} &=& \frac{1}{\sqrt{2}}\langle h^{+}_i  b_{i\uparrow}
   h^{+}_j  b_{j\downarrow} - h^{+}_i  b_{i \downarrow}
   h^{+}_j  b_{j \uparrow}\rangle
   \nonumber \\
   &\simeq &
    \langle h^{+}_i h^{+}_j \rangle
   \frac{1}{\sqrt{2}} \langle  b_{i\uparrow}
    b_{j\downarrow} - b_{i \downarrow}
   b_{j \uparrow}\rangle \equiv F^{+}_{ij}\, \varphi_{ij},
\label{m4a}\\
   b_{ij} &=& \frac{1}{\sqrt{2}} ( h^{+}_i  b_{i\uparrow}
   h^{+}_j  b_{j\downarrow} - h^{+}_i  b_{i \downarrow}
   h^{+}_j  b_{j \uparrow}) \simeq h^{+}_i   h^{+}_j
   \varphi_{ij}.
\label{m4}
\end{eqnarray}
Within these approximations  we obtain the effective Hamiltonian
for holons
\begin{eqnarray}
H_{h} & \simeq & - \sum_{i \neq j} \, {\tilde t_{ij}}\,h_i\,
h^{+}_j\,  - \mu \sum_{i }\,h_i\, h^{+}_i
\nonumber \\
&-& \frac{1}{2} \sum_{i \neq j}{\tilde J_{ij}} \,
 (  \langle h_j\, h_i \rangle \, h^{+}_i \, h^{+}_j
+  h_j \, h_i \langle h^{+}_i \, h^{+}_j\rangle ) ,
 \label{m5}
\end{eqnarray}
where the effective hopping parameter and the exchange energy are
given by
\begin{eqnarray}
{\tilde t_{ij}} &=& t_{ij}\, \langle  b_{i\uparrow}^+\,
b_{j\uparrow} + b_{i \downarrow}^+\,
   b_{j \downarrow}\rangle = t_{ij}\, \langle  S_{i}^{+}\,
S_{j}^{-} +  S_{i}^{-}\, S_{j}^{+} \rangle
   \nonumber \\
  {\tilde J_{ij}} &=& J_{ij}\, |\varphi_{ij}|^2 =
  J_{ij}\,\frac{1}{2} \, | \langle  b_{i\uparrow}\,
    b_{j\downarrow} - b_{i \downarrow}\,
   b_{j \uparrow}\rangle|^2 \, .
 \label{m6}
\end{eqnarray}
 To obtain a phase diagram for the RVB order parameter
$\,B_{ij}=F_{ij}^{+} \, \varphi_{ij}\,$ as a function of
temperature $T$ and hole doping $\delta $ one should solve a
self-consistent system of equations for the both order parameters,
holon $\, F_{ij}\,$ and spinon $\, \varphi_{ij} \,$ ones. Here we
consider only equations for the holon order parameter by
suggesting that there exists  a region of $\,(T, \, \delta) \,$
where the spinon order parameter is nonzero.

Introducing ${\bf k}$-vector representation for the correlation
functions:
\begin{eqnarray}
F_{ij}& = &\langle h_i \, h_j \rangle = \frac{1}{N}\sum_{{\bf k}}
\, {\rm e}^{i({\bf k(i-j))}} \,F({\bf k})
 \nonumber \\
&=& \frac{i}{N}\sum_{{\bf k}} \, \sin({\bf k(i-j)}) \,\langle
h_{\bf k}\, h_{\bf - k} \rangle \, ,
\label{m9a}   \\
\varphi_{ij} \   & = & \frac{1}{\sqrt{2}} \, \langle
b_{i\uparrow}\, b_{j\downarrow} - b_{i \downarrow}\,
   b_{j \uparrow}\rangle
 = \frac{1}{N}\sum_{{\bf k}} \, {\rm e}^{i({\bf k(i-j))}} \,
 \varphi({\bf k})
 \nonumber \\
   &=&\frac{2i}{N}\sum_{{\bf k}}
   \, \sin({\bf k(i-j)})\frac{1}{\sqrt{2}} \, \langle b_{{\bf k}\uparrow}\,
 b_{{\bf - k}\downarrow} \rangle \, ,
\label{m9}
\end{eqnarray}
the Hamiltonian~(\ref{m5}) can be written in  ${\bf k}$-space as
\begin{eqnarray}
H_{h} & = &  \sum_{\bf k} \,(\varepsilon({\bf k})-\mu_h ) \,h_{\bf
k}^{+}\, h_{\bf k}\,
\nonumber \\
& -& \frac{1}{2} \sum_{\bf k}\, \{ \Delta({\bf k}) \, h^{+}_{\bf
-k} \, h^{+}_{\bf k}  +
 \Delta^{+}({\bf k}) \,  h_{\bf k} \, h_{- \bf k}  \}\, ,
 \label{m10}
\end{eqnarray}
where the chemical potential for holons $\, \mu_h = -\mu \,$ and
the holon spectrum $\,\varepsilon({\bf k})$ according to
Eq.~(\ref{m6}) is written in the form (cf. Eq.~(\ref{32a})):
\begin{eqnarray}
\varepsilon({\bf k})&=& {\tilde t}({\bf k}) = 4t \,\gamma({\bf
k})\,2 \langle S_{i}^+\,S_{i+a_1}^{-} \rangle
 \nonumber \\
 &+& 4t' \,\gamma'({\bf k})\,2 \langle  S_{i}^+\,S_{i+a_2}^{-}\rangle ,
\label{m11}
\end{eqnarray}
Here we assumed that the spin correlation functions have $s$-wave
symmetry, e.i., $\,\chi_{i,i\pm a_{x}} = \chi_{i,i \pm a_{y}}\,$.
 We also introduced the holon gap function  defined
by the equation
\begin{equation}
\Delta({\bf k})= - \Delta({\bf -k}) = \frac{1}{N } \sum_{{\bf q}}
 {\tilde J}({\bf k-q})\, \langle h_{\bf q}\, h_{\bf - q}
\rangle .
 \label{m13}
\end{equation}
In Eqs.~(\ref{m11}),~(\ref{m13}) we used notation
$\,\varepsilon({\bf k}), \, \Delta({\bf k})\, $ for the holon
spectrum and the holon gap to distinguish them from that ones  of
physical electrons  in Sec.~2.2.

 The normal and anomalous correlation functions can be easily
calculated for the BCS-type Hamiltonian~(\ref{m10}):
\begin{eqnarray}
 \langle h^{+}_{\bf q}\, h_{\bf q} \rangle & = &\frac{1}{2}\,
\left( 1- \frac{\varepsilon({\bf q})-\mu_h}{ E({\bf q})}
\tanh\frac{E({\bf q})}{2T} \right),  \label{m14a}\\
  F({\bf q})& =& \langle h_{\bf q}\,h_{\bf - q} \rangle =
 \frac{\Delta({\bf q})}{2 E({\bf q})} \tanh\frac{E({\bf q})}{2T} ,
 \label{m14}
\end{eqnarray}
where the quasipartical spectrum
 $\, E({\bf k}) = [(\varepsilon({\bf k})-\mu_h)^{2}+
 |\Delta({\bf  k})|^2]^{1/2} \,$.
From these equations follow the self-consistent equations for the
gap function  and the average number of holes $\delta$ which
defines the holon chemical potential $\mu_{h}$:
\begin{eqnarray}
\Delta({\bf k}) &=& \frac{1}{N } \sum_{{\bf q}}
 {\tilde J}({\bf k-q})\,
\frac{\Delta({\bf q})}{2 E({\bf q})} \tanh\frac{E({\bf q})}{2T},
\label{m15}\\
 \delta &=& \frac{1}{N } \sum_{{\bf q}}\,
 \langle h^{+}_{\bf q}\, h_{\bf q} \rangle
 \nonumber \\
 &=&\frac{1}{N } \sum_{{\bf
q}}\frac{1}{2}\left(1- \frac{ \varepsilon({\bf q})-\mu_h}{E({\bf
q})}\right) \, \tanh\frac{E({\bf q})}{2T}.
 \label{m16}
\end{eqnarray}
These equations are identical to the results obtained
in~\cite{Kochetov00} where  the path-integral representation for
the $t$-$J$ model was applied and  an effective BCS-type
Hamiltonian analogous to Eq.~(\ref{m10}) was derived. Starting
from the MFA for RVB order parameter as in Eq.~(\ref{m2}), the
authors introduced MFA for the spinon auxiliary field that
resulted in the spin-charge separation as in our Eq.~(\ref{m4}).
Therefore, we have proved that the results of the path-integral
representation for the $t$-$J$ employed in Ref.~\cite{Kochetov00}
are equivalent to the MFA for the slave fermion - hard-core boson
approach considered in this section.

\subsection{Holon Green functions}

To avoid uncontrollable approximation caused by the decoupling of
the Hubbard operators on the same lattice site in Eq.~(\ref{m3a})
used in  MFA for RVB state in the Hamiltonian~(\ref{m2}), in the
present section we consider projection technique for the holon GF
for the spinon-holon model. By using the spinless fermion
hard-core boson representation~(\ref{s1}) we write the Hamiltonian
of the $t$-$J$ model~(\ref{9}) as follows:
\begin{eqnarray}H_{t-J} & = & - \sum_{i \neq j} \,  t_{ij}\,h_i\,
h^{+}_j\,( S_{i}^{+}\, S_{j}^{-} +  S_{i}^{-}\, S_{j}^{+} )
 \nonumber \\
 &-& \mu \sum_{i }\,h_i\, h^{+}_i - \frac{1}{4} \sum_{i \neq j} \, J_{ij} \;
  h_i\, h_j \, h^{+}_j \, h^{+}_i \;
\nonumber \\
 & \times&
( S_{i}^{+}\, S_{j}^{-} -  S_{i}^{-}\, S_{j}^{+} )( S_{j}^{+}\,
S_{i}^{-} -  S_{j}^{-}\, S_{i}^{+} ).
 \label{g1}
\end{eqnarray}
To obtain equation for the holon GF we apply the projection
technique described in Sec.~2.1. By introducing the matrix GF for
holon operators:
\[\hat{G}_{i j}^{h} (t-t') =
 \left( \begin{array}{cc}
    \langle\langle h_i(t) | h^+_j (t') \rangle\rangle &
    \langle\langle h_i (t) | h_j (t') \rangle\rangle \\
   \langle\langle h^+_i (t) | h^+_j (t') \rangle\rangle  &
   \langle\langle h^+_i(t)  | h_j (t') \rangle\rangle
  \end{array} \right) \, ,
\]
we can obtain equation of motion for
the GF as discussed in Sec.~2.1 (see Eqs.~(\ref{12})--(\ref{23})).
For the zero-order GF in MFA we get the following result:
\begin{eqnarray}
\hat G^{(h, 0)} ({\bf k},\omega)= \frac{\omega
\hat\tau_{0}+(\tilde\varepsilon ({\bf k}) - \mu_h)
 \hat \tau_{3}+{\tilde \Delta} ({\bf k}) \hat \tau_{1} }
 {\omega^{2}-(\tilde \varepsilon ({\bf k}) - \mu_h)^{2}-
 \mid \tilde \Delta ({\bf k})\mid^{2} },
 \label{g3}
\end{eqnarray}
where the the holon dispersion and the gap function are given by
the equations:
\begin{eqnarray}
\tilde \varepsilon({\bf k})&= &{\tilde t}({\bf k})-
\frac{1}{N}\sum_{{\bf q}}\, \{ {\tilde J'}({\bf k-q})-{\tilde
J'}(0) \}\, \langle h_{\bf q}^{+}\, h_{\bf q} \rangle ,
\label{g4}\\
\tilde\Delta({\bf k})& = & \frac{1}{N } \sum_{{\bf q}}
 {\tilde J'}({\bf k-q})\, \langle h_{\bf q}\, h_{\bf - q}
\rangle .
 \label{g5}
\end{eqnarray}
In comparison with MFA for RVB state considered previously, in the
GF approach we obtain additional renormalization for the holon
dispersion, Eq.~(\ref{m11}), while the gap equation has the same
form, Eq.~(\ref{m13}), but with a different renormalized exchange
interaction:
\begin{eqnarray}\label{g6}
{\tilde J'}_{ij} &=& J_{ij}\,\frac{1}{2} \langle | S_{i}^{+}\,
S_{j}^{-} - S_{i}^{-}\, S_{j}^{+} |^2 \rangle \nonumber \\
 &=&
J_{ij}\,( \frac{1}{4}- \langle S_{i}^{z} \, S_{j}^{z}\rangle ).
\end{eqnarray}
By using equation of motion method instead of MFA for RVB state we
managed to take into account spinon correlations in the effective
holon interaction~(\ref{g6}) and obtained a simple formula for it
by using the identities: $\,S_{i}^{+}\, S_{i}^{+}=0, \,
S_{i}^{-}\, S_{i}^{+}= (1/2) - S_{i}^{z} \,$ in the last equation.
From Eq.~(\ref{g6}) we get the following  estimation for the
effective exchange energy for the nearest neighbors: $\, \tilde J'
\simeq (0.5 \div 0.25)\, J \,$ if we assume the AFM Ne\'{e}l
phase: $\,\langle S_{i}^{z} \, S_{j}^{z}\rangle = - 1/4$ or
completely neglect AFM correlations:  $\, \langle S_{i}^{z} \,
S_{j}^{z}\rangle = 0\,$,  respectively.

\section{Results and discussion}

\subsection{Spectral density}

Spectral density for physical electron excitations in the lower
Hubbard subband within the $t$-$J$ model which is defined by the
equation
\begin{eqnarray}\label{r1}
A_{\sigma}({\bf k}, \omega) = - \frac{1}{\pi}{\rm Im}
\langle\langle X_{{\bf k}}^{0 \sigma } \, |\, X_{{\bf k}}^{
\sigma 0} \rangle\rangle_{\omega +i\epsilon},
\end{eqnarray}
satisfies the sum rule:
\begin{eqnarray}\label{r2}
\sum_{\sigma}\, \int\limits\sb{-\infty}\sp{+\infty}
  {\rm d}\omega \, A_{\sigma}({\bf k}, \omega)&=&
\sum_{\sigma}\,\langle \{X_{i}^{0\sigma},\, X_{i}^{\sigma
0}\}\rangle \nonumber \\
 &=& 2\, Q =1 + \delta .
\end{eqnarray}
In the MFA for the GF in terms of the HO, Eq.~(\ref{30}), we get:
\begin{equation}\label{r3}
A_{\sigma}({\bf k}, \omega) = Q\, u_{\bf k} \,\delta(\omega -
E_{{\bf k}})\, + Q\, v_{\bf k} \,\delta(\omega + E_{{\bf k}}),
\end{equation}
where
\begin{equation}\label{r3a}
   u_{\bf k} = \frac{1}{2}(1+\frac{(\varepsilon_{\bf k}
-\tilde
 \mu)}{ E_{\bf k}}) , \quad
  v_{\bf k} = \frac{1}{2}(1-\frac{(\varepsilon_{\bf k} -\tilde
 \mu)}{ E_{\bf k}}).
 \end{equation}
 In MFA the spectral density~(\ref{r3}) satisfies the sum
rule~(\ref{r2}) but have no incoherent background which appears
if one takes into account the self-energy corrections as shown in
Ref.~\cite{Plakida99}.

In the spinon-holon representation~(\ref{s1}) the spectral
density, as follows from Eq.~(\ref{s5d}), does not obey the sum
rule~(\ref{r2}):
\begin{eqnarray}
&&\sum_{\sigma}\, \int\limits\sb{-\infty}\sp{+\infty}
  {\rm d}\omega \,A_{\sigma}^{(sh)}({\bf k}, \omega) =
\nonumber \\
&=&
 \sum_{\sigma}\,  \int\limits\sb{-\infty}\sp{+\infty}
  {\rm d}\omega  \,  \{- \frac{1}{\pi} {\rm Im}
\langle\langle  h^{+}_i \, b_{i\sigma}\, | h^{}_i \,
b_{i\sigma}^{+}  \rangle\rangle_{\omega +i\epsilon}\,\}
 \nonumber \\& = &
  \sum_{\sigma}\, \langle\{  h^{+}_i \, b_{i\sigma}\, ,
h^{}_i \, b_{i\sigma}^{+} \}\rangle   = 1.
   \label{r4}
\end{eqnarray}
 MFA for the spinon-holon GF in Eq.~(\ref{r4}) results in
the spin-charge separation  which defines the spectral
density~(\ref{r4}) as a convolution of the anticommutator holon
and the commutator spinon GF~\cite{Feng97}:
\begin{eqnarray}
 A_{\sigma}^{(sh)}({\bf k}, \omega)
 &= & \frac{1}{N}\sum_{q}
 \frac{1}{2 \pi\sp{2}}\!
  \int\sb{-\infty}\sp{+\infty}
   \int\sb{-\infty}\sp{+\infty}\!
 \frac{{\rm d}\omega\sb{1} {\rm d}\omega\sb{2}}
 {\omega - \omega\sb{1} - \omega\sb{2}}
  \nonumber \\
  & \times & [ \tanh (\omega\sb{1}/ 2T ) + \coth (\omega\sb{2}/2T)]
  \nonumber \\
& \times &
  \mbox{Im} \langle\!\langle h^+_{\bf q} \!\mid\!
        h_{\bf q}  \rangle\!\rangle\sb{\omega\sb{1}} \;
  \mbox{Im}\langle\!\langle S^{\bar \sigma}_{\bf k-q}\!\mid\!
        S^{\sigma}_{\bf k-q}\rangle\!\rangle\sb{\omega\sb{2}}.
\label{r5}
\end{eqnarray}
 In this representation the spectral density of physical electron
 excitations  is given by a superposition of the spectral density
for spinless fermion (holon) excitations and a background produced
by  spin excitations~\cite{Feng97}. In MFA the holon spectral
density in the paired state below $T_c$ is defined by the
GF~(\ref{g3}):
\begin{eqnarray}\label{r6}
A^{(h)}({\bf k}, \omega) &=& - \frac{1}{\pi}{\rm Im}
\langle\langle h_{{\bf k}}^{+}\, |\, h_{{\bf k}}
\rangle\rangle_{\omega +i\epsilon} \nonumber\\
 &= &\tilde u_{\bf
k} \,\delta(\omega + \tilde E({\bf k}))
 + \tilde v_{\bf k} \,\delta(\omega - \tilde E({\bf k})),
\end{eqnarray}
where  $\, \tilde E({\bf k}) = [(\tilde \varepsilon ({\bf k}) -
\mu_h)^2 + \mid \tilde \Delta ({\bf k})\mid ^2 ]^{1/2}\,$ is the
quasiparticle energy and
 $\,\tilde u_{\bf k}, \,\tilde v_{\bf k} \,$ are given by Eq.~(\ref{r3a})
but with quasiparticle energies for holons.   The spectral density
of electron excitations is measured in the angle-resolved
photoemission (ARPES) experiments which provide  information
concerning the symmetry of the gap function. As discussed below,
the symmetry of superconducting gap function $\,\Delta_{\bf k}\,$
defined by Eq.~(\ref{35}) appears to be $d$-wave, while  the holon
gap function $\, \Delta ({\bf k})\,$ in Eq.~(\ref{m15}) has odd
symmetry of the $p$-type as in the triplet pairing. This $p$-wave
gap symmetry has never observed in ARPES experiments in cuprates.


\subsection{Gap symmetry}
 For models with strong electron correlations
the $s$-wave component of  superconducting gap must be  strongly
suppressed due to on-site Coulomb correlations.  For the $t$-$J$
model it follows from  the constraint of no double occupancy on a
single site given by the identity:
\begin{equation}
\langle \hat{c}_{i, \sigma} \hat{c}_{i,-\sigma}\rangle =
\frac{1}{N} \sum_{\bf k} \langle \hat{c}_{{\bf k}, \sigma}
\hat{c}_{-{\bf k}, -\sigma}\rangle = 0 ,
 \label{r7}
\end{equation}
for the physical  electron operators
$\hat{c}_{i,\sigma}=c_{i,\sigma}(1-n_{i,-\sigma})$. Since the
anomalous correlation function $\langle \hat{c}_{{\bf k}, \sigma}
\hat{c}_{ {\bf -k}, -\sigma}\rangle$ is proportional to the gap function
$\Delta({\bf k})$, Eq.~(\ref{r7}) imposes a certain constraint on
the symmetry of the gap function. In particular, for the
solution~(\ref{33a}) in  MFA   Eq.~(\ref{r7})
reads~\cite{Plakida89,Yushankhai91}:
\begin{eqnarray}
\langle X_{i}^{0 {\sigma}}  X_{i}^{0 \bar\sigma} \rangle &=&
\frac{1}{N} \sum_{\bf q}\, \langle X_{{\bf q}}^{0 {\sigma}}
X_{{\bf -q}}^{0 \bar\sigma} \rangle \nonumber\\
 &=& \frac{Q}{N} \,
\sum_{\bf q}\, \frac{\Delta_{{\bf q}}^{\sigma}}{2 E_{\bf q}}
\tanh\frac{E_{\bf q}}{2T} =0 \, .
 \label{r8}
 \end{eqnarray}
For a tetragonal lattice  the Fermi surface (FS) is invariant
under the $C_4$-axis rotation in the ${\bf k}$-space and therefore
to satisfy the condition~(\ref{r8}) for $\, E_{\bf q} > 0 \,$ the
gap function $\, \Delta_{{\bf q}}^{\sigma}\,$ should change its
sign along the FS. It means that the symmetric $s$-wave solution,
$\, \Delta_{s}(k_{x},k_{y}) \propto (\cos q_x + \cos q_y)\,$, does
not fit Eq.~(\ref{r8}), while the  $d$-wave solution with $B_{1g}$
or $B_{2g}$ symmetry, $\Delta_{d}(k_{x},k_{y})=  -
\Delta_{d}(k_{y},k_{x})$, satisfy the condition~(\ref{r8}). In
general, Eq.~(\ref{r8}) should be considered as a constraint on
the symmetry of the gap function in the superconducting phase and
solutions which violate this constraint should be disregarded.

Now we consider the symmetry of the spinon and holon order
parameters. The condition~(\ref{r7}) for the RVB order parameter
reads
\begin{equation}
  B_{ii} = \frac{1}{N}\sum_{\bf k}\, B({\bf k}) =
 \frac{1}{N} \sum_{\bf k, q}\, \varphi({\bf k-q})\, F^{+}({\bf q})
 = 0 ,
 \label{r9}
\end{equation}
which  also imposes a constraint on the symmetry of the spinon and
holon order parameters.
 Since the symmetry of the holon  pairing order parameter, Eq.~(\ref{m9a}),
 for spinless fermions is odd, imposed by their anticommutation relations,
 it results in the odd  symmetry of the gap $\Delta({\bf k})$ in
Eq.~(\ref{m13}). For a tetragonal lattice the symmetry is given by
the odd two-dimensional irreducible representation $E_u$ which can
be modelled by the function:  $\, \Delta({\bf k})  \propto
\,\eta_p^{\pm}({\bf k})=  (\sin k_x \pm \sin k_y)\,$ as in the
 $p$-wave triplet pairing. The same holds for the spinon order
 parameter in Eq.~(\ref{m9}),
 $\, \varphi({\bf k})\propto \, \eta_p^{\pm}({\bf k})\,$.

The  RVB order parameter $\, B_{ij} = \langle b_{ij} \rangle =
\varphi_{ij}F_{ij}^{+} \,$ as a product of two antisymmetric
order parameters with $E_u$ symmetry has  either $A$ ($s$-wave)
or $B_{1g}$ ($d$-wave) symmetry. Namely, if we adopt only the
nearest neighbor pairing for  both the order parameters:
\begin{eqnarray}
\varphi_{ij} &= &\varphi_{i, i+a_{x}}\{
 (\delta_{j, i+a_{x}}-\delta_{j,i-a_{x}})   \pm \,
 (\delta_{j,i+a_{y}} - \delta_{j,i-a_{y}})\} , \nonumber\\
 F_{ij}^+ &= & F_{i, i+a_{x}}^+
\{ (\delta_{j, i+a_{x}}-\delta_{j,i-a_{x}})   \pm \,
 (\delta_{j,i+a_{y}} - \delta_{j,i-a_{y}})\},
  \nonumber
\end{eqnarray}
then for the Fourier transform of the RVB order parameter we get
\begin{eqnarray}\label{r11}
 B({\bf k})& = &\sum_{\bf  q}\, \varphi({\bf k-q})\, F^{+}({\bf q})
= 2 b\, \sum_{\bf  q}\, \{ \sin (k_x -q_x)
\nonumber \\
 & \pm &\sin(k_y - q_y)\}
 \,\{ \sin q_x \pm \sin q_y \} \nonumber \\
&=& - b\,(\cos k_x \pm \cos k_y)
\end{eqnarray}
where $\, b= \varphi_{i, i+a_{x}}\,F_{i, i+a_{x}}^+ \,$ and the
sign $+ \, (-)$ in the last line corresponds to the same
(different) signs in the first line. The solution~(\ref{r11})
satisfies the condition~(\ref{r9}) for both the $s$- and $d$-wave
symmetry: $\sum_{\bf k} B({\bf k}) \propto
  \sum_{\bf k}(\cos k_x \pm \cos k_y) = 0$.

However,  the explicit solution for the holon order
parameter~(\ref{m14}) shows  its more complicated ${\bf
k}$-dependence which proves that the frequently used model for the
nearest neighbor pairing  is inadequate. For the
solution~(\ref{m14}) the condition~(\ref{r9}) for the RVB order
parameter reads:
\begin{eqnarray}\label{r12}
&& \sum_{\bf k, q}\, \varphi({\bf k-q})\, F^{+}({\bf q})=
\nonumber\\
 & = &
 \sum_{\bf k, q}\, \varphi({\bf k-q}) \,
 \frac{\Delta^{+}({\bf q})}{2 E({\bf q})} \tanh\frac{E({\bf
 q})}{2T} =0 ,
\end{eqnarray}
which also  imposes a requirement on the symmetry of the singlet
(RVB) order parameter which should have   only  the $d$-wave
symmetry, $\, B(k_x, k_y) = - B(k_y, k_x) \,$,  to satisfy this
condition.

It is interesting to compare results for the gap equation
 derived in MFA for  the $t$-$J$ model and that one for the Hubbard model.
By applying the projection technique for the GF for the Hubbard
model one  can get the following equation for the gap function
(see, e.g.,~\cite{Plakida01}):
\begin{equation}
 \Delta_{ij\sigma} \propto \langle X\sb{i}\sp{02} N\sb{j} \rangle
  = \langle c\sb{i\downarrow}c\sb{i\uparrow} N\sb{j}\rangle,
\label{r13}
\end{equation}
where we have used the identity for the Hubbard operators, $\,
X\sb{i}\sp{02} = X\sb{i}\sp{0\downarrow} X\sb{i}\sp{\downarrow 2}
=   c\sb{i\downarrow}c\sb{i\uparrow} \, $. From Eq.~(\ref{r13})
follows that the pairing occurs on one lattice site but in the
different Hubbard subbands. By using equation of motion for the GF
$\, \langle \langle X\sb{i}\sp{02} (t) \mid N\sb{j} (t') \rangle
\rangle\,$ the anomalous correlation function $\,\langle
X\sb{i}\sp{02} N\sb{j} \rangle \,$ can be calculated without any
decoupling that results in the same gap equation as in the MFA for
the $t$-$J$ model~\cite{Plakida01}: $\,\Delta_{ij\sigma} =
J_{ij}\, \langle X_{i}^{0\sigma} X_{j}^{0\bar\sigma} \rangle /
Q\,$ where we have used the notation of the present paper.
Therefore the symmetry constraint considered above is also
applicable for the gap solutions obtained in the Hubbard model.

\subsection{Superconducting $\bf T_c$}

In the present section we compare the pairing temperature $T_c$
defined by the gap equation for physical electrons~(\ref{35}) and
that one for holons, Eq.~(\ref{m15}). We study $T_c$ as a function
of corresponding chemical potential $\,\nu = {\mu}/{W}\,$ for a
given  value of the  pairing interaction $\,\lambda =  {J}/{W}\,$
where all energies are measured in units of the renormalized half
bandwidth for pairing fermions $\,W= 4\, \tilde t\,$:
\begin{equation}
 e({\bf k})= \frac {\varepsilon({\bf k})}{W } = \gamma({\bf
k})+ \tau \gamma^{'}({\bf k}),   \quad d({\bf k}) = \frac
{\Delta(\bf k)}{W} ,
 \label{t1}
\end{equation}
where $\, \tau = {t'_{eff}}/{t_{eff}}\,$. The gap
Eqs.~(\ref{35}),~(\ref{m15}) (or ~(\ref{g5})) can be written in
the form
\begin{eqnarray}
d({\bf k}) &=& \frac{4\lambda}{N } \sum_{{\bf q}} \,
\frac{\gamma({\bf k-q}) \, d({\bf q})}{2[(e({\bf q})-\nu)^2 +
d({\bf q})]^{1/2}}
 \nonumber \\
 & \times & \tanh\frac{[(e({\bf q})-\nu)^2 + d({\bf q})]^{1/2}}{2 T},
 \label{t2}
\end{eqnarray}
Let us consider $T_c(\nu)$ for different pairing symmetry, $
d({\bf k})= d_{\alpha}\,\eta_{\alpha}({\bf k}) $:
\begin{eqnarray}
s-{\rm wave:} \quad \eta_s({\bf k})= (\cos k_x + \cos k_y),
\nonumber \\
 d-{\rm wave:} \quad \eta_d({\bf k})= (\cos k_x -\cos k_y),
\nonumber \\
 p-{\rm wave:} \quad \eta_p^{\pm}({\bf k})=(\sin k_x \pm \sin k_y).
\label{t3}
\end{eqnarray}
By integrating Eq.~(\ref{t2}) with the corresponding symmetry
parameter $\eta_{\alpha}({\bf k})$ we obtain Eq.~(\ref{t2}) for
$T=T_c$ in the same form for any symmetry:
\begin{eqnarray}
 \frac{1}{\lambda}& = & \frac{1}{N} \sum_{{\bf k}}\, ( \eta_{\alpha}({\bf k}) )^2
\frac{1}{2 (e({\bf k}) - \nu )} \tanh \frac{e({\bf k})-\nu}
{2T_c^{(\alpha)}}
\nonumber \\
 & = & \frac{1}{2} \int_{-1}^{+1} \frac {d\epsilon} {\epsilon -\nu}
N_{\alpha}(\epsilon) \tanh \frac{\epsilon-\nu}{2T_c^{(\alpha)}} \,
, \label{t4}
\end{eqnarray}
 if we introduce an effective  density of state (DOS) for the
corresponding symmetry, $\, \alpha = s, d, p $:
\begin{equation}
N_{\alpha}(\epsilon)= \frac{1}{N} \sum_{{\bf k}}\, (
\eta_{\alpha}({\bf k}) )^2 \, \delta(\epsilon - e({\bf k}))\, ,
\label{t5}
\end{equation}
which is normalized
\[
\int_{-1}^{+1}\, d\epsilon \, N_{\alpha}(\epsilon)=1 ,\quad
 {\rm since} \quad
 \frac{1}{N} \sum_{{\bf k}}\, (
\eta_{\alpha}({\bf k}) )^2 =1.
\]
The results of calculation of  the effective DOS, Eq.~(\ref{t5}),
for different symmetry $\, \alpha = s, d, p \,$ is presented on
Fig.~1 for $t' =0$. From  this dependence it is easy to draw a
conclusion that the $T_c(\nu)$ function will follow the dependence
$\,N_{\alpha}(\nu)\,$ for corresponding symmetry since the
effective coupling constant $ V \simeq \lambda N_{\alpha}(\nu)$
reaches its maximum value at the maximum value of $
N_{\alpha}(\nu)$. In the logarithmic approximation a solution for
$T_c(\mu)$, Eq.~(\ref{t4}), can be written in the conventional BCS
form:
\begin{equation}
 T_c \simeq  \sqrt{\mu (W-\mu)} {\exp}{(-1/V)} , \quad
V \simeq J \, N_{\mu}  , \label{tc}
\end{equation}
but with the prefactor proportional to the Fermi energy, $\mu =
E_F$ which can result in high-$T_c$. The highest $T_c(\nu)$
appears for the $d$-wave pairing and the lowest $T_c(\nu)$ -- for
the $p$- wave pairing since in the former case the van Hove
singularity gives a strong contribution for DOS ($\,[\eta_d({\bf k}
= (\pm \pi,0))]^2 = 4\,$), while in the latter case it is
completely suppressed ($\,[\eta_p({\bf k} = (\pm \pi,0))] = 0\,$).
This estimation can be checked by a direct numerical solution of
Eq.~(\ref{t4}) (see, e.g.~\cite{Yushankhai91,Plakida99} and
~\cite{Kuzmin99} for the $p$-wave symmetry).
\begin{figure}[t]
\centering
{\epsfig{file=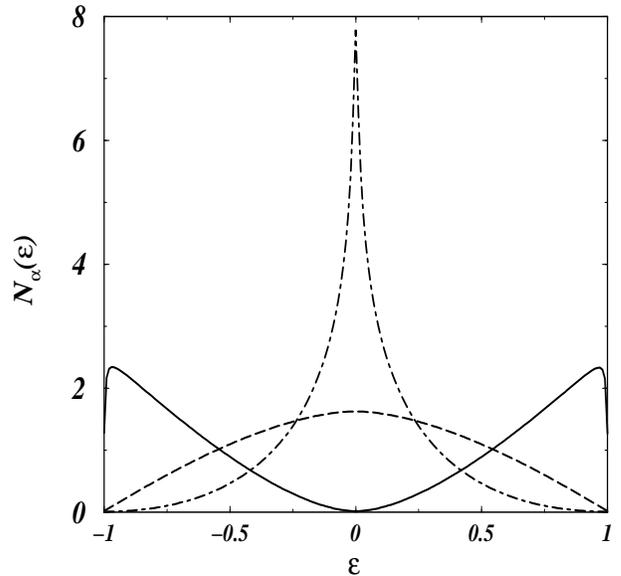,height=80mm,width=80mm,angle=-90}}
\vspace*{1mm}
 \caption{Effective density of
states, Eq.~(\ref{t5}), for $s$ (solid line), $d$ (dot-dashed
line) and $p$ (dashed line) symmetry.}
 \end{figure}

\section{Conclusion}

In the present paper we consider superconducting pairing mediated
by the exchange interaction which is generic for system with
strong electron correlations as cuprates. The mechanism of the
exchange pairing is  the lowering of kinetic energy of electron
pairs due to their coherent hopping  between different Hubbard
subbands. Since the excitation energy of this hopping  is much
larger then the Fermi energy the retardation effects in the
exchange interaction are negligible~\cite{Plakida01} which results
in pairing of all electrons (holes) in the conduction band and a
high-$T_c$ proportional to the Fermi energy, Eq.~(\ref{tc}).

To obtain estimation for the superconducting $T_c$ it is tempting
to use a mean-field approximation for the exchange interaction
within the $t$-$J$ model. However, meaningful physical results can
be obtained only if one takes into account strong electron
correlations on a rigorous basis which is provided by the Hubbard
operator  technique. Any auxiliary field representations applied
within MFA inevitably violates rigorous commutation relations for
HO which may result in unphysical conclusions. In the present
paper we have proved this by comparing the results for
superconducting gap equation derived  within the  slave fermion -
hard-core boson representation, Eq.~(\ref{s1}), and HO technique
for the Green functions (Sec.~2.1). In the former method the
projected character of physical electron operators is neglected
that results in double counting of empty states and violation of
the sum rule, Eq.~(\ref{r4}). The spin-charge separation which
occurs in MFA, Eqs.~(\ref{m4a}),~(\ref{m4}), results in  separate
equations for two order parameters for spinons and holons instead
of one equation for physical electrons as in the GF method. The
gap equation~(\ref{m15}) for spinless fermions has only
antisymmetric solutions  which results in the  $p$-wave gap for
the quasiparticle excitations in superconducting state never
observed in ARPES experiments in cuprates and much lower $T_c$
then for the $d$-wave pairing given by Eq.~(\ref{35}) for physical
electrons.

The obtained results within MFA for the spinless fermion -- hard
core bosons in Sec.~2 appears to be identical to the path-integral
representation for the $t$-$J$ model employed in
Ref.~\cite{Kochetov00}. Therefore the latter results have the same
flaws as discussed above which casts doubts on  approaches based
on the idea of spin-charge separation. Violation of the local
constraint in the two-band Hubbard model may lead also to
unphysical results, as shown in Ref.~\cite{Zhang93}.\\

{\bf Acknowledgments}\\

The author thanks V. Oudovenko for  preparing Fig.~1 and
V.Yu.~Yushankhai for discussions. The author is grateful to Prof.~P.~Fulde for
the hospitality extended to him during his stay at MPIPKS, where the final
version of the paper was prepared.


\begin{thebibliography}{25}
\bibitem{exchange} Sometimes  a coefficient $\, (1/2) \,$
instead of $\, (1/4) ,$ is written  in Eq.~(\ref{9}) for the
exchange energy $J$~(see, e.g.,~\cite{Plakida89,Yushankhai91}).
\end{thebibliography}
\end{document}